\input{aipcheck}

\documentclass[
    ,final            % use final for the camera ready runs
  ]
  {aipproc}

\layoutstyle{8x11double}

\begin{document}

\title{Phase diagram and critical properties in the Polyakov--Nambu--Jona-Lasinio model
}


\author{C. A. de Sousa}{
  address={Centro de Física Computacional, Departamento de Física,
   Universidade de Coimbra, P-3004-516 Coimbra, Portugal}
}

\author{P. Costa}{
  address={Centro de Física Computacional, Departamento de Física,
   Universidade de Coimbra, P-3004-516 Coimbra, Portugal}
}

\author{M. C. Ruivo}{
  address={Centro de Física Computacional, Departamento de Física,
 Universidade de Coimbra, P-3004-516 Coimbra, Portugal}
}

\author{H. Hansen}{
  address={IPNL, Universit\'e de Lyon/Universit\'e Lyon 1, CNRS/IN2P3, 4 rue E.Fermi, F-69622
Villeurbanne Cedex, France} }

%%%%%%%%%%%%%%%%%%%%%%%%%%%%%%%%%%%%%%%%%%%%
%% ABSTRACT
%%%%%%%%%%%%%%%%%%%%%%%%%%%%%%%%%%%%%%%%%%%%

\begin{abstract}
 We investigate the phase diagram of the so-called Polyakov--Nambu--Jona-Lasinio model at
finite temperature and nonzero chemical potential with three quark flavours. Chiral and
deconfinement phase transitions are discussed, and the relevant order-like parameters are
analyzed. The results are compared with simple thermodynamic expectations and lattice
data. A special attention is  payed  to the critical end point: as the strength of the
flavour-mixing interaction becomes weaker, the critical end point moves to low
temperatures and can even disappear.

\end{abstract}

\maketitle

%%%%%%%%%%%%%%%%%%%%%%%%%%%%%%%%%%%%%%%%%%%%
%% MAINMATTER
%%%%%%%%%%%%%%%%%%%%%%%%%%%%%%%%%%%%%%%%%%%%

There is strong evidence that quantum chromodynamics (QCD) is the fundamental theory of
strong interactions. Its basic constituents are quarks and gluons that are confined in
hadronic matter. It is believed that at high temperatures and densities hadronic
matter should undergo a phase transition into a new state of matter, the  quark-gluon
plasma (QGP). A challenge of theoretical  studies based on QCD is to  predict the
equation of state, the critical end point (CEP) and the nature of the phase transition.

As the equation of state of strong interacting matter given by lattice simulations is now
considered  as a function of temperature and a limited range of chemical potential,
QCD-like models, as Nambu--Jona-Lasinio (NJL) type models, are suited to provide guidance
and information relevant to observe experimental signs of deconfinement and QGP features.

NJL type models, that take into account only quark degrees of freedom, give the correct
chiral properties; static gluonic degrees of freedom are then introduced in the NJL
Lagrangian through an effective gluon potential in terms of Polyakov loop with the aim of
including features of both chiral symmetry breaking and deconfinement.

Hence our calculations are performed in the framework of an extended SU(3)$_f$ PNJL
Lagrangian, which includes the 't Hooft instanton induced interaction term that breaks
the U$_A$(1) symmetry, and the quarks are coupled to the (spatially constant) temporal
background gauge field $\Phi$ \cite{Fukushima,Ratti}. The
Lagrangian reads:
%
%%%%%%
\begin{eqnarray}\label{eq:lag} {\mathcal L}&=& \bar q(i \gamma^\mu D_\mu-\hat m)q +
\frac{1}{2}\,g_S\,\,\sum_{a=0}^8\, [\,{(\,\bar q\,\lambda^a\, q\,)}^2\, \nonumber\\&+&
{(\,\bar q \,i\,\gamma_5\,\lambda^a\, q\,)}^2\,] \,+\, g_D\,\{\mbox{det}\,[\bar
q\,(1+\gamma_5)\,q] \nonumber\\ &+&\mbox{det} \,[\bar q\,(1-\gamma_5)\,q]\}
%\nonumber\\&=&
- \mathcal{U}\left(\Phi[A],\bar\Phi[A];T\right).
\end{eqnarray}
%%%%%%

The covariant derivative is defined as $D^{\mu}=\partial^\mu-i A^\mu$, with
$A^\mu=\delta^{\mu}_{0}A_0$ (Polyakov gauge); in Euclidean notation $A_0 = -iA_4$.  The
strong coupling constant $g$ is absorbed in the definition of $A^\mu(x) = g {\cal
A}^\mu_a(x)\frac{\lambda_a}{2}$, where ${\cal A}^\mu_a$ is the (SU(3)$_c$) gauge field
and $\lambda_a$ are the (color) Gell-Mann matrices.

The  effective potential for the (complex) field $\Phi$ adopted in our parametrization of
the PNJL model  reads:
%%%%%%
\begin{eqnarray}
    \frac{\mathcal{U}\left(\Phi,\bar\Phi;T\right)}{T^4}
    &=&-\frac{a\left(T\right)}{2}\bar\Phi \Phi +
    b(T)\mbox{ln}[1-6\bar\Phi \Phi  \nonumber\\&+&4(\bar\Phi^3+ \Phi^3)-3(\bar\Phi
    \Phi)^2],
    \label{Ueff}
\end{eqnarray}
%%%%%%
where
%%%%%%
\begin{equation}
    a\left(T\right)=a_0+a_1\left(\frac{T_0}{T}\right)+a_2\left(\frac{T_0}{T}
  \right)^2\,\mbox{ and }\,\,b(T)=b_3\left(\frac{T_0}{T}\right)^3.
\end{equation}
%%%%%

The effective potential exhibits the feature of a phase transition from color confinement
($T<T_0$, { the minimum of the effective potential being at $\Phi=0$}) to color
deconfinement ($T>T_0$, the minima of the effective potential occurring at $\Phi \neq
0$).

%%%%%%%%%%%%%%%%%%%%%%%%%%%%%%%%%%%%
\begin{figure}[ptb]
%\begin{center}
\hspace{-0.25cm}\includegraphics[width=0.37\textwidth]{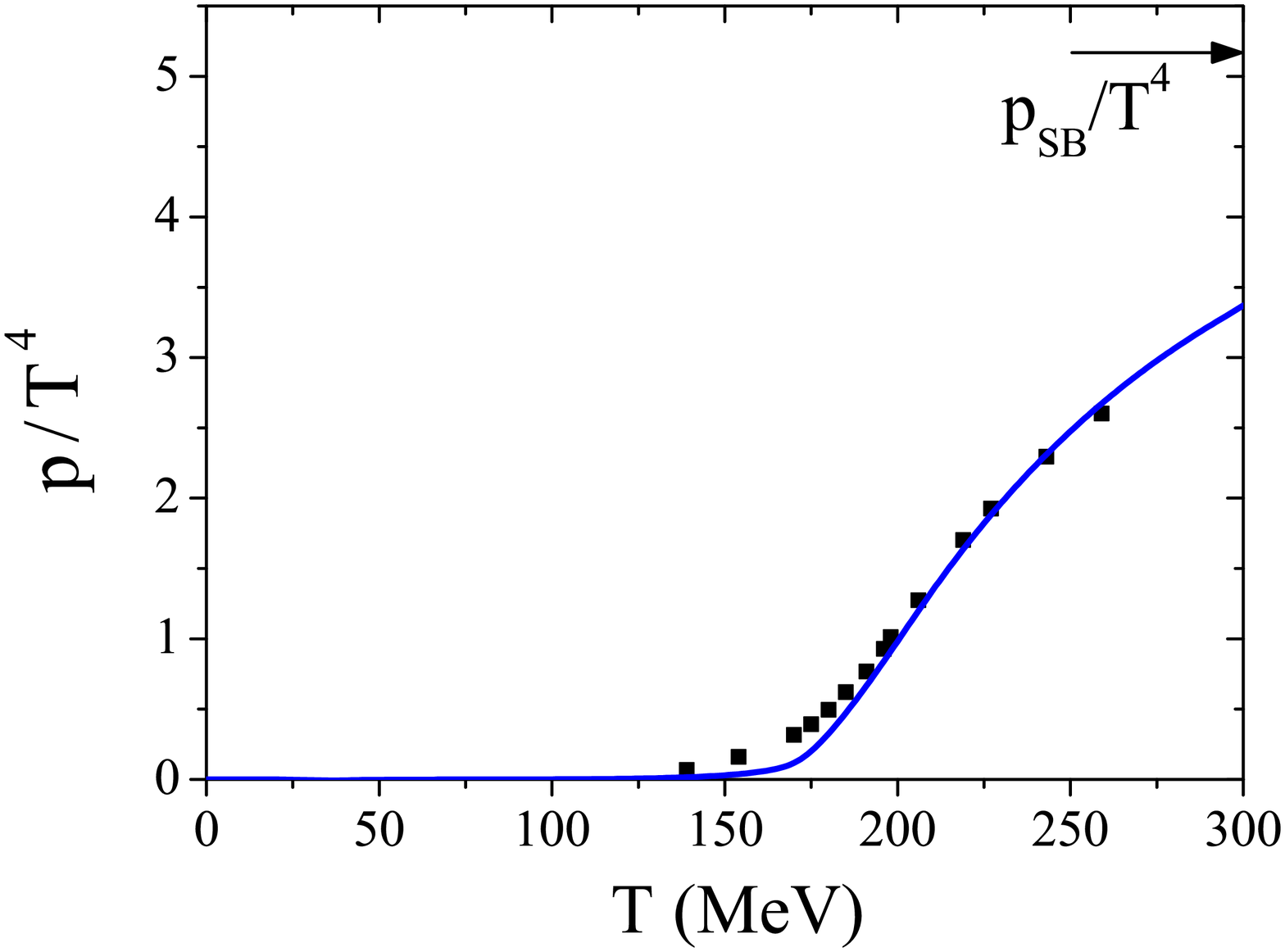}
\hspace{-0.75cm}\includegraphics[width=0.37\textwidth]{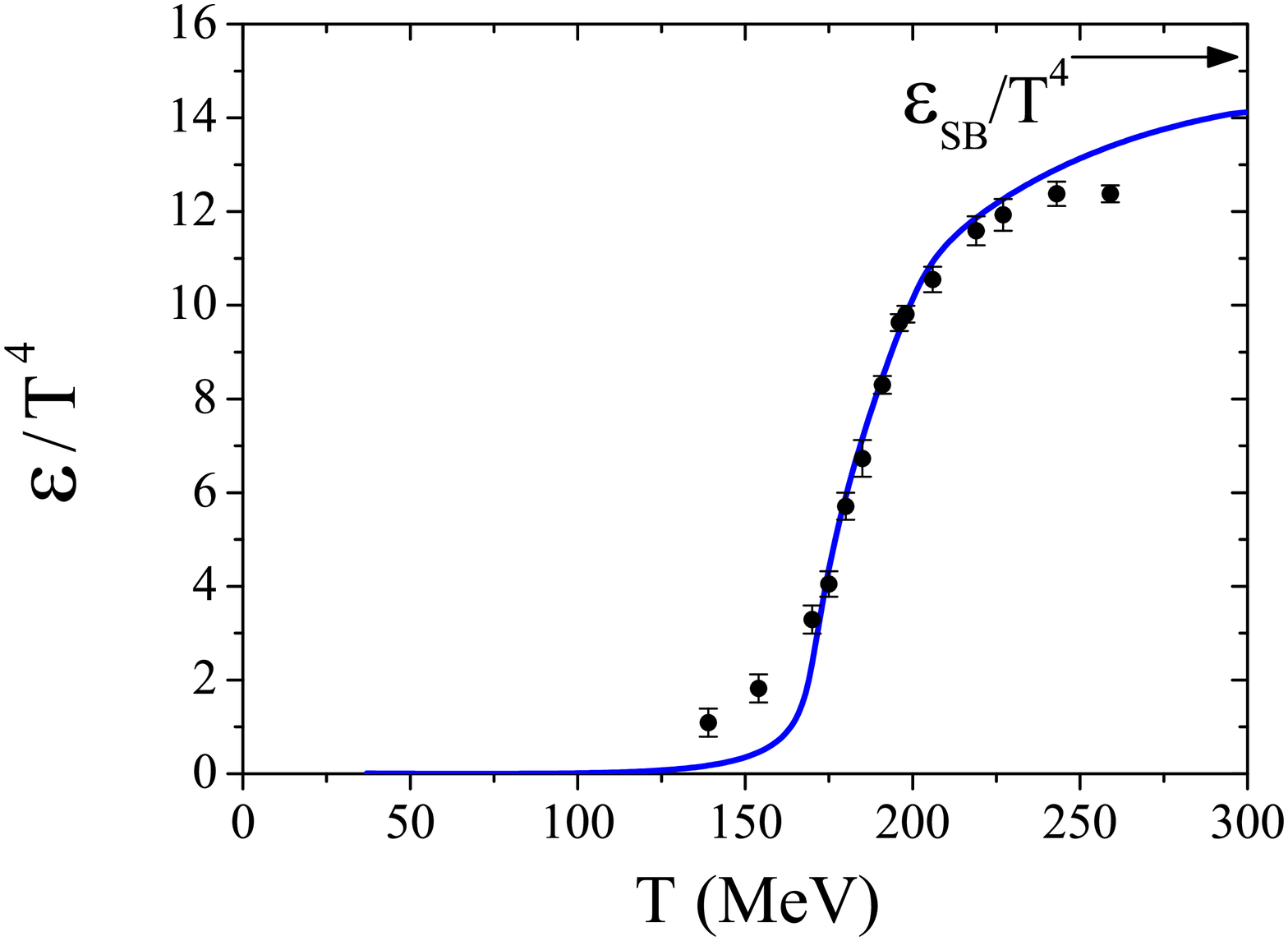}
\hspace{-0.75cm}\includegraphics[width=0.37\textwidth]{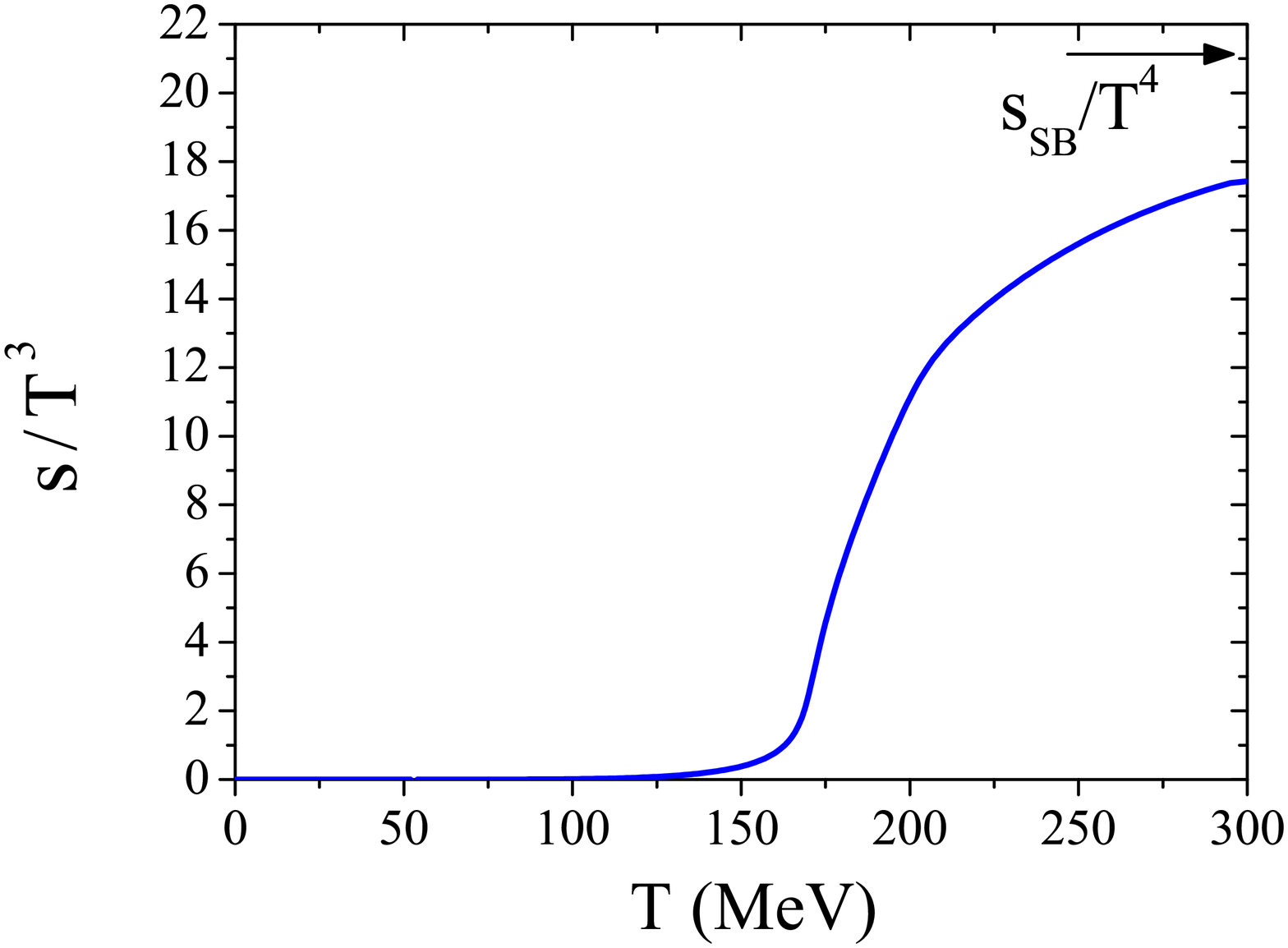}
%\end{center}
\caption{\leftskip=1cm \rightskip=1cm Scaled pressure $(p)$,  energy  per particle
$(\epsilon)$, and entropy $(s)$ as a function of the temperature at zero chemical
potential. The data points are taken from \cite{lattice}. }
\end{figure}
%%%%%%%%%%%%%%%%%%%%%%%%%%%%%%%%%%%%

The parameters of the effective potential $\mathcal{U}$ are given by $a_0=3.51$, $a_1=
-2.47$, $a_2=15.2$ and $b_3=-1.75$. These parameters have been fixed in order to
reproduce the lattice data for the expectation value of the Polyakov loop and QCD
thermodynamics in the pure gauge sector.
The parameter $T_0$,  the critical temperature for the deconfinement phase transition
within a pure gauge approach, was fixed to $270$ MeV, according to lattice findings.
 This choice ensures an almost exact coincidence between chiral crossover and deconfinement at
zero chemical potential, as observed in lattice calculations.
We notice, however,  that a rescaling of $T_0$ to $210$ MeV may be needed in some cases
in order to get agreement between model calculations and thermodynamic quantities
obtained on the lattice (see Fig. 1). Let us stress that this modification of $T_0$
 is essentially done for rescaling reasons (an absolute temperature scale has not a very
strong meaning in these kind of Ginzburg--Landau model based on symmetry) but does not
change drastically the physics.

The parameters of the NJL sector are: $m_u = m_d = 5.5$~MeV,
$m_s = 140.7$ MeV, $g_S\Lambda^2 = 3.67$, $g_D \Lambda^5 = -12.36$
and $\Lambda = 602.3$ MeV, which are fixed to
reproduce the values of the coupling constant of the pion, $f_\pi\,=\,92.4$ MeV, and the
meson masses of the pion, the kaon, the $\eta$ and $\eta^\prime$, respectively,
$M_\pi\,=\,135$ MeV, $M_K\,=\,497.7$ MeV, $M_\eta\,=\,514.8$ MeV and
$M_{\eta^\prime}\,=\,960.8$ MeV.

%%%%%%%%%%%%%%%%%%%%%%
\vskip0.2cm
%%%%%%%%%%%%%%%%%%%%%%
In the limit of vanishing quark chemical potential, significant information
on the phase structure of QCD at high temperature is obtained from lattice calculations.

In Fig. 1, we plot the scaled pressure, the energy and the entropy as functions of the
temperature compared with recent lattice results (see Ref. \cite{lattice}).
Since the transition to the high temperature phase is a rapid crossover rather than a
phase transition, the pressure, the entropy and the energy densities are continuous
functions of the temperature. We observe a similar behavior in the three curves: a sharp
increase in the vicinity of the transition temperature.

Asymptotically, the QCD pressure for $N_f$ massless
quarks and $(N_c^2 - 1)$ massless gluons is given  ($\mu=0$) by:
%%%%%%%
\begin{equation}\label{pSB}
\frac{p_{SB}}{T^4}\,=\,(N_c^2 - 1)\,\frac{\pi^2}{45}\,+\,N_c\,N_f\,\frac{7\,\pi^2}{180},
\end{equation}
%%%%%%%
where the first term denotes the gluonic contribution and the second term the fermionic
one. The results follow the expected tendency and go to the free gas values (or
Stefan--Boltzmann (SB) limit).
The inclusion of the Polyakov loop effective potential ${\cal U}(\Phi,\bar\Phi;T)$ (it
can be seen as an effective pressure term mimicking the gluonic degrees of freedom of
QCD) is required to get the correct limit (indeed in the NJL model the ideal gas limit is
far to be reached due to the lack of gluonic degrees of freedom).

Some comments are in order concerning the role of our regularization procedure for $T>T_c$,
that allows for the presence of high momentum quark states \cite{varios}.
In this temperature range, due to the presence of such states, the physical
situation is dominated by the significant decrease of the constituent quark masses.
This allows for an ideal gas behavior of almost massless quarks
with the correct number of degrees of freedom.

Let us notice that, just below $T_c$, the pressure and the energy fail to reproduce the
lattice points: for example there is a small underestimation of the pressure and energy
in the model calculations. It was pointed out that the lack of mesonic correlations in the PNJL
model is responsible for, at least, a fraction of this discrepancy.

%%%%%%%%%%%%%%%%%%%%%%
\vskip0.2cm
%%%%%%%%%%%%%%%%%%%%%%

%%%%%%%%%%%%%%%%%%%%%%%%%%%%%%%%%%%%
\begin{figure}[ptb]
%\begin{center}
\hspace{-0.25cm}\includegraphics[width=0.50\textwidth]{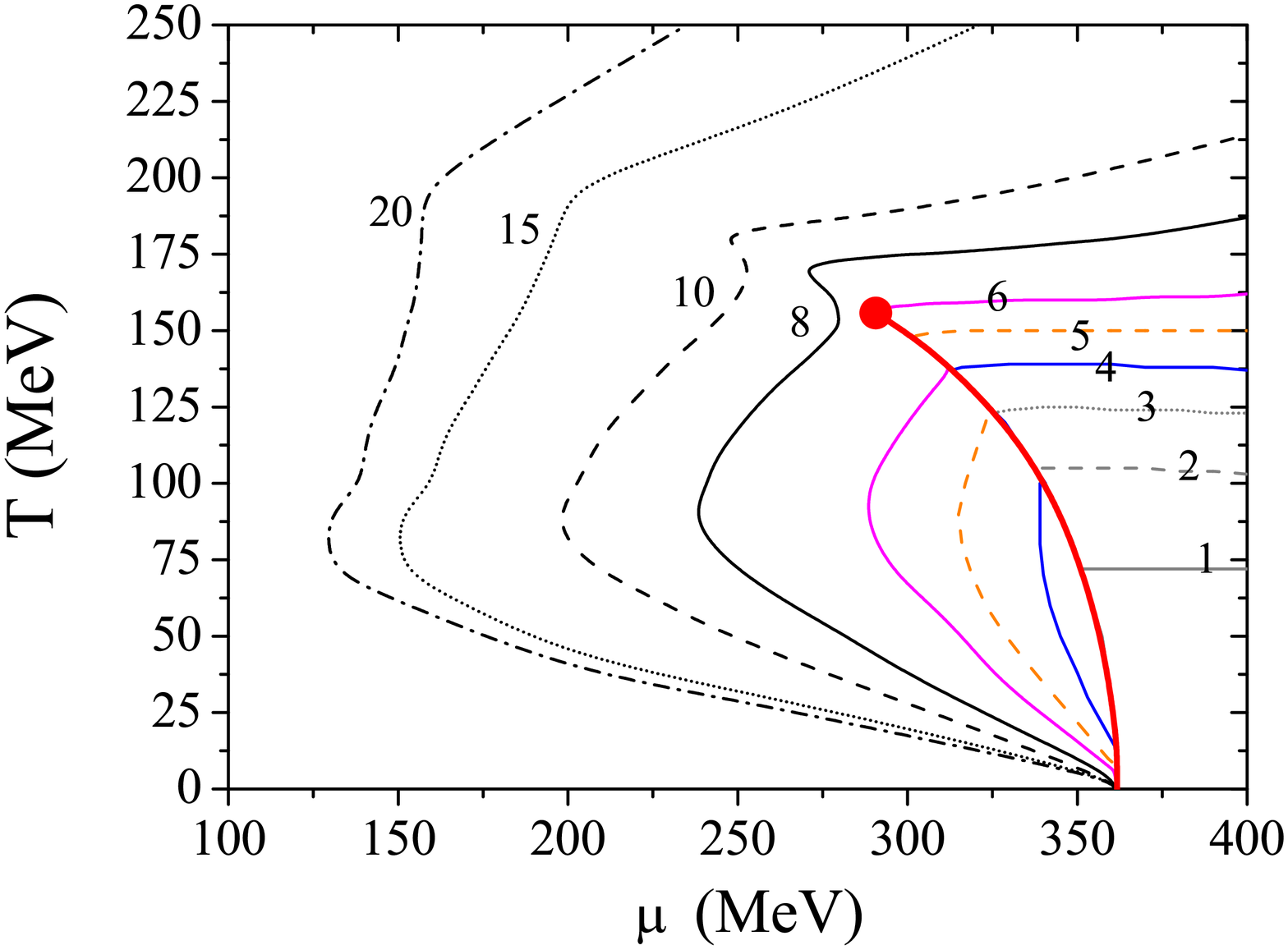}
\hspace{-0.75cm}\includegraphics[width=0.50\textwidth]{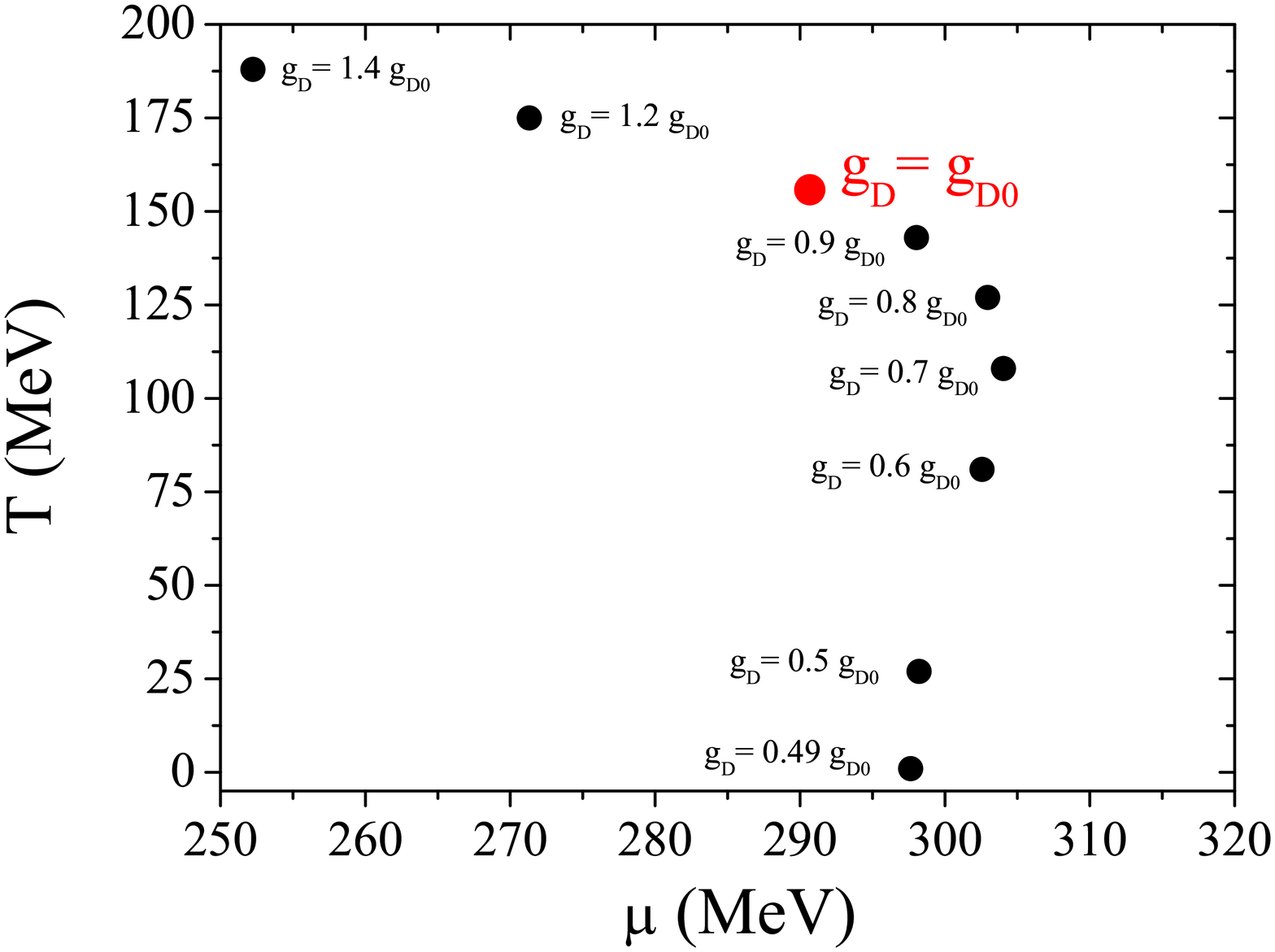}
%\hspace{-0.75cm}\includegraphics[width=0.37\textwidth]{entrop_Temp.eps}
%\end{center}
\caption{Left: Isentropic trajectories in the $(T,\mu)$ plane.  Right: Dependence of
the location of the CEP on the strength of the t Hooft coupling constant $g_D$.}
\end{figure}
%%%%%%%%%%%%%%%%%%%%%%%%%%%%%%%%%%%%

Our numerical results for the isentropic lines in the $(T,\mu)$ plane are shown in
Fig. 2 (left side). We start the discussion by analyzing the behavior of the isentropic
lines in the limit $T\rightarrow 0$. Our convenient choice of the model parameters allows
a better description of the first order transition than other treatments of the NJL
(PNJL) model. This choice is crucial to obtain important results: the criterion of
stability of the quark droplets is fulfilled, and, in
addition, simple thermodynamic expectations in the limit $T\rightarrow 0$ are verified.
In fact, in this limit $s \rightarrow 0$ according to the third law of thermodynamics
and, as $\rho_B \rightarrow 0$ too, the satisfaction of the condition
$s/\rho_B\,=\,const.$ is insured.

We observe that the isentropic lines with $s/\rho_B=1,...,6$ come from the region of
symmetry partially restored and attain directly the phase transition.
Consequently, all isentropic trajectories directly terminate in the end of the first
order transition line at $T=0$.
The trajectories with  $s/\rho_B>6$ go directly to the crossover region and display a
smooth behavior, although those that pass in the neighborhood of the CEP show a slightly
kink behavior \cite{varios}.

%%%%%%%%%%%%%%%%%%%%%%%%%%%%%%%%%%%%%%%%%%%%%

The location and even the existence of the CEP in the phase diagram is a matter of
debate. While different lattice calculations predict the existence of a CEP \cite{Fodor},
the absence of the CEP in the phase diagram was seen in recent lattice QCD results
\cite{deForcrand}, where the first order phase transition region near $\mu=0$ shrinks in
the quark mass and $\mu$ space when $\mu$ is increased \cite{deForcrand}. Due to the
importance of the  U$_A$(1) anomaly  and its influence on several observables, it is
demanding to investigate possible changes in the location of the CEP in the $(T,\,\mu)$
plane when the anomaly strength is modified. In Fig. 2 (right side) we show the location
of the CEP for several values of $g_D$ compared to the results for $g_{D_0}$, the value
used for the vacuum. As already pointed out by K. Fukushima in \cite{Fukushima}, we also
observe that the location of the CEP depends on the value of $g_D$. In fact,  our results
show that the existence or not of the CEP is determined by the strength of the anomaly
coupling, the CEP getting closer to the $\mu$ axis as $g_D$ decreases.

%%%%%%%%%%%%%%%%%%%%%%%%%%%%%%%%%%%%%%%%%%%%
\vskip 0.2cm
%%%%%%%%%%%%%%%%%%%%%%%%%%%%%%%%%%%%%%%%%%%%

We investigated the phase diagram of the so-called PNJL model at finite temperature and
nonzero chemical potential with three quark flavours. Chiral and deconfinement phase
transitions are discussed, and the relevant order-like parameters are analyzed. The
results are compared with simple thermodynamic expectations and lattice data. A special
attention is  payed  to the critical end point: as the strength of the flavour-mixing
interaction becomes weaker, the critical end point moves to low temperatures and can even
disappear.

The sets of parameters  used  is compatible with the formation of stable droplets at zero
temperature, insuring  the satisfaction of important thermodynamic expectation like the
Nernst principle.
Consequently,  all the trajectories  directly terminate in the same point of the
horizontal axes at $T=0$.
The picture provided here is a natural result in these type of quark models with no
change in the number of degrees of freedom of the system in the two phases.

Other important role is played by the regularization procedure which, by allowing high
momentum quark states, is essential to obtain the required increase of extensive
thermodynamic quantities, insuring the convergence to the SB limit of
QCD. In this context the  gluonic degrees of freedom also play a special role.

The successful comparison with lattice results shows that the model calculation provides
a convenient tool to obtain information for  systems from zero to nonzero chemical
potential which is of particular importance to the knowledge of the equation of state of
hot and dense matter.

\vskip0.3cm
%%%%%%%%%%%%%%%%%%%%%%%%%%%%%%%%%%%%%%%%%%%%%%%%%%%%%%%%%%%%
%\section*{Acknowledgements}
Work supported by  Centro de F\'{\i}sica Computacional and F.C.T. under Project
\linebreak No. CERN/FP/83644/2008.

\vspace{-0.5cm}

%%%%%%%%%%%%%%%%%%%%%%%%%%%%%%%%%%%%%%%%%%%%%%%%
%% BACKMATTER
%%%%%%%%%%%%%%%%%%%%%%%%%%%%%%%%%%%%%%%%%%%%%%%%

%\begin{theacknowledgments}
 % ............
%\end{theacknowledgments}



\bibliographystyle{aipproc}   % if natbib is available
\bibliographystyle{aipproc}   % if natbib is available

%\bibliography{sample}

\IfFileExists{\jobname.bbl}{}
 {\typeout{}
  \typeout{******************************************}
  \typeout{** Please run "bibtex \jobname" to optain}
  \typeout{** the bibliography and then re-run LaTeX}
  \typeout{** twice to fix the references!}
  \typeout{******************************************}
  \typeout{}
 }

\end{document}